\documentclass{PoS}

\title{Propagation and stability of relativistic jets}

\ShortTitle{Propagation and stability of relativistic jets}

\author{\speaker{Manel Perucho}\thanks{A footnote may follow.}\\
        Departament d'Astronomia i Astrof\'{\i}sica, Universitat de Val\`encia, C/ Dr. Moliner, 50, 46100, Burjassot, Valencian Country, Spain.\\
        Observatori Astron\`omic, Universitat de Val\`encia, C/ Catedr\`atic Jos\'e Beltr\'an 2, 46980, Paterna, Valencian Country, Spain.
        E-mail: \email{manel.perucho@valencia.edu}}

\abstract{A simple look at the steady high-energy Universe reveals a clear correlation with outflows generated around compact objects (winds and jets). In the case of relativistic jets, they are thought to be produced as a consequence of the extraction of rotational energy from a Kerr black hole (Blandford-Znajek), or from the disc (Blandford-Payne). A fraction of the large energy budget provided by accretion and/or black hole rotational energy is invested into jet formation. After formation, the acceleration and collimation of these outflows allow them to propagate to large distances away from the compact object. The synchrotron cooling times demand that re-acceleration of particles takes place along the jets to explain high-energy and very-high-energy emission from kiloparsec scales. At these scales, jets in radio galaxies are divided in two main morphological/luminosity types, namely, Fanaroff-Riley type I and II (FRI, FRII), the latter being more luminous, collimated and edge-brightened than the former, which show clear hints of decollimation and deceleration. In this contribution, I summarise a set of mechanisms that may contribute to dissipate magnetic and kinetic energy: Magnetohydrodynamic instabilities or jet-obstacle interactions trigger shocks, shearing and mixing, which are plausible scenarios for particle acceleration. I also derive an expression for the expected distance in which the entrainment by stellar winds starts to be relevant, which is applicable to FRI jets. Finally, I discuss the differences in the evolutionary scenarios and the main dissipative mechanisms that take place in extragalactic and microquasar jets.}

\FullConference{High Energy Phenomena in Relativistic Outflows VII - HEPRO VII\\
		9-12 July 2019\\
		Facultat de F\'{\i}sica, Universitat de Barcelona, Spain}

\begin{document}

\section{Introduction}

 Relativistic outflows carry large amounts of energy ($10^{43}-10^{47}\,{\rm erg/s}$ in the case of extragalactic jets and $10^{36}-10^{40}\,{\rm erg/s}$ in the case of microquasars) from the environment of compact objects, where they are triggered, to large distances (hundreds of kpc for extragalactic jets and parsecs for microquasars). The jets are formed by extraction of rotational energy from a Kerr black-hole \cite{bz77,tch11,bk12,bkp12} or from an accretion disc \cite{bp82}. A part of that energy is released in the form of radiative output, throughout the whole electromagnetic spectrum, mainly via synchrotron and inverse-Compton (either external or internal) processes. The radiative losses are typically taken as a negligible part of the total energetic budget, so jets are modelled as adiabatic. Nevertheless, this can be incorrect in regions in which the magnetic field is intense and radiative losses are strong, e.g., close to the jet formation region. The adiabatic approximation is probably valid once the jet is mass-loaded and the intensity of the magnetic field has dropped due to possible investment in acceleration after formation and jet expansion. At kiloparsec scales, the jets show a morphological dichotomy between collimated, edge-brightened FRII radiogalaxies and less collimated, dimmer FRI radiogalaxies \cite{fr74}.
 
    Jets can be modelled as plasmas, because the Larmor radius of particles moving around magnetic field lines is very small compared to the scales of the problem \cite{br74}, which can be understood as the magnetic field giving consistency to the flow. The jets carry their energy flux in the form of kinetic, internal and magnetic energy flux. The relative weight of the three energetic {\emph channels} determines the way in which the jet interacts with its environment, the instabilities that may develop and the possibility of particle acceleration at shocks.  
 
  The cooling times of energetic particles are inversely proportional to the particle Lorentz factor and to the square of the magnetic field. In terms of the critical emitting frequency, it can be written as (see, e.g., \cite{ba16}):
  
 \begin{equation}
 t_c \, = \, 3 \sqrt{3 \pi} \sqrt{\frac{m_e c e}{\nu_c}} \frac{1}{\sigma_T \beta^2} B^{-3/2},
 \end{equation} 
where $m_e$ and $e$ are the electron mass and charge, respectively, $c$ is the speed of light, $\sigma_T$ is the Thomson cross-section, $\beta$ is the particle speed with respect to the speed of light, $\nu_c$ is the critical emitting frequency, and $B$ is the magnetic field in Gauss. Using a relatively low observing frequency, typically used to image large-scale jet structures, 178\,MHz, we obtain:
  
 \begin{equation}
 t_c \, \simeq 1.2 \times 10^8 B^{-3/2}\, {\rm s}.
 \end{equation}  
 
   If the bulk velocity is close to the speed of light, this translates in a distance of $\simeq 4$ light-years for 1 Gauss, or $1.2\times 10^5\,{\rm ly}$ for $1 \,{\rm mG}$. Taking into account that the magnetic field drops with distance (by simple expansion and conservation of the magnetic flux), we can infer that little particle re-acceleration is needed. However, the cooling times and distances shorten with $\nu_c^{-1/2}$, and, therefore, X-ray emission would be unexpected at kiloparsec scales. Still, X-rays are observed in FRI radio galaxies within the deceleration region (a few hundred pc to several kpc) and even at larger scales in FRII jets. Moreover, extended gamma-ray emission has been reported by \emph{Fermi} observations of a close FRI radio galaxy (Centaurus A). These observations imply particle acceleration mechanisms at work at large distances from the active nucleus. Interestingly, it has been noted that a larger percentage of gamma-ray detections among FRI jets than FRII's with the \emph{Fermi} observatory, which cannot be accounted for in terms of source populations \cite{dg16,gr12}. This is counter-intuitive taking into account that FRIs are weaker than FRIIs in terms of jet energy budgets (see, e.g., \cite{gc01}). 
   
   Possible acceleration processes include Fermi-I type acceleration process at shocks, acceleration in turbulent flows, shear acceleration at the transition between the jet and its surrounding medium, or magnetic reconnection (see \cite{rl18} and references therein). In this contribution I summarise a set of possible magnetohydrodynamics scenarios that can take place in jets and become the frame in which either kinetic or magnetic energy can be dissipated, increasing the internal energy budget and accelerating particles to non-thermal energies. A more extended revision of the subject can be also found in \cite{pe19}. This contribution is structured as follows: In Section 2, I summarize the possible instabilities that can develop in jets and contribute to jet deceleration and energy dissipation. In Section 3, I describe the scenario of jet-obstacle interaction and the role of such events on jet evolution. In Section 4, I derive an expression for the distance in which we expect the mass load by stars to become relevant (this is also the distance in which we expect a stronger dissipation generated by collisions). I also discuss the differences and similarities in energy dissipation scenarios between microquasar and extragalactic jets. Finally, in Section 5 I present the conclusions that can be extracted from this contribution. 
    
\section{Jet instabilities}

 When jets are formed they are probably magnetically dominated, and, thus, sub-Alfv\'enic. The magnetic field structure generated by rotation is presumably toroidal. Acceleration may stretch the lines, but the conservation of the magnetic flux favours that toroidal field dominates the magnetic field structure with distance, in the case of expanding outflows. Farther downstream, the investment of internal and magnetic energy in the process of bulk acceleration, and entrainment, change the jet energy flux to particle dominated. These changes are relevant to the instability modes that may develop.   
 
   The linear growth of instabilities is studied by linearizing the RMHD equations and introducing wave-like perturbations (see \cite{pe19} and references therein). The solutions include complex values for the frequency and wavenumber, where the imaginary part becomes the growth-rate of the instability. Numerical simulations can then be used to study the post-linear and non-linear regimes, although high numerical resolution is demanded to follow the growth of linear modes from small amplitudes \cite{pe04}.
 
  While the jet is magnetically dominated, the current-driven instability (CDI) may develop (see \cite{bo13,kim17,kim18,bo19}). This instability is favoured mainly by toroidal configurations. From an observational point of view, there are no hints of global disruption at parsec scales, which may have, at least, two different explanations: 1) the instability modes do not develop fast enough to affect the flow before the flow conditions change, or 2) there is no global magnetic field configuration and the instability only develops at small scales within the jet flow, if at all. Among the stabilising mechanisms for the CDI, a strong poloidal field component, jet expansion, wide shear-layers or winds shielding the jet, or a non-negligible azimuthal velocity component (see also \cite{ma16}) have been shown to delay the growth of disrupting CDI modes.

  As pointed out, a drop in the magnetic energy flux is expected during the acceleration phase, which, together with the flow acceleration itself, bring the jet to becoming super-Alfv\'enic. At this point, the CDI is suppressed with respect to the Kelvin-Helmholtz instability (KHI) modes (see, e.g., \cite{ha07}). As opposed to the case of the sub-parsec scales, a number of observed jet structures have been related to the growth of KHI modes at parsec-scales (e.g., \cite{lz01,har03,har05,har11,vg19}). Still, no global jet disruption on the base of the development of such modes has been reported, but in one case (S5~0836+710), in which the jet decollimation and deceleration has been attributed to the growth of a helical mode out to kiloparsec scales \cite{pe12a,pe12b,ka19}.
  
  Again, because the growth-rates of global instabilities are related to the time the waves need to cross the jet and bounce to its boundaries (see, e.g., \cite{pc85,pk15}), jet expansion \cite{har86} and axial velocity (Lorentz factor, \cite{pe05,pe10}) reduce them and favour jet stability. In the case of the Lorentz factor, it contributes to stretch the distance between bounces, so even if the jet opening angle is small, the growth-times or distances may be long. 
   
  Although the nonlinear development of KHI and CDI modes has been related to the deceleration of FRI jets \cite{pm07,ro08,tb16}, the morphological dichotomy between FRI and FRII radiogalaxies seems to be related to the growth of small-scale instabilities that trigger mixing from the jet boundaries to its axis \cite{lb14}. In this context, short wavelength KHI modes \cite{pe10}, Rayleigh-Taylor instability (RTI, \cite{mm13,ma17,to17,mk07,mk09}) or the centrifugal instability (CFI, \cite{gk18a,gk18b}) have been suggested to develop in expanding and recollimating jets. Nevertheless, although FRI jets have relatively large opening angles at the decelerating region, there is no observational hint of such large-scale recollimation shocks in archetypical FRI jets, so the solution to the problem does not seem to be unique. 
  
  Even if the aforementioned list of possible developing instabilities plays no evident role in the FR dichotomy, they may certainly contribute to dissipate part of the magnetic and/or kinetic energy and thus contribute to long term deceleration. The dissipation induced by instabilities is relatively small during the linear phase of the amplitude growth, and it is associated to the oscillations generated, but it becomes relevant in the cases in which the amplitudes reach post-linear or non-linear values. In this case, the dissipation takes place via shocks and turbulent mixing.

\section{Interactions}

  Stars and clouds are numerous in galactic cores and can penetrate the jet as they orbit around the nucleus. This process triggers a strong bow-shaped shock wave around the obstacle \cite{ko94}. In the case of clouds, the shock crosses them, heating the gas, which expands and enlarges the interaction cross-section, whereas in the case of stars, the stellar wind equilibrates the jet flow at a distance that is determined by setting the wind-to-jet momentum ratio to unity:
  
\begin{equation}\label{eq:rs}
R_s=\sqrt{\frac{\dot{M}_{\rm w}\,v_{\rm w}} {4 \pi\,\rho_{\rm j}\,\gamma_{\rm j}^2\,v_{\rm j}^2 }},
\end{equation}  
 where $R_s$ is the distance between the contact discontinuity and the star, $\dot{M}_{\rm w}$ is the stellar wind mass flux, $v_{\rm w}$ is the wind velocity, and the subscript ${\rm j}$ indicates jet parameters: rest-mass density, $\rho$, Lorentz factor, $\gamma$, and velocity, $v_{\rm j}$. This expression can also be written as
   
\begin{eqnarray}\label{eq:rs2}
R_s=2.14 \times 10^{12} \left(\frac{\dot{M}_{\rm w}}{10^{-11}\,{\rm M_\odot \, yr^{-1}}}\right)^{1/2} \,\left(\frac{v_{\rm w}}{10\,{\rm km s^{-1}}}\right)^{1/2} \times  \, \nonumber \\ \left(\frac{L_{\rm j}}{10^{43}\,{\rm erg s^{-1}}} \right)^{-1/2}  \,\left(\frac{v_{\rm j}}{c}\right)^{-1/2} \,\left(\frac{R_{\rm j}}{1\,{\rm pc}}\right)^{1/2} \,\left(\frac{h_{\rm j}}{c^2}\right)^{1/2}\, {\rm cm},
\end{eqnarray}  
where we have used that $\rho_{\rm j}\,\gamma_{\rm j}^2\,v_{\rm j}^2 \,=\, L_{\rm j}\, v_{\rm j} / (\pi R_{\rm j}^2 h_{\rm j})$, with $L_{\rm j}$ the jet power, $R_{\rm j}$ the jet radius and $h_{\rm j}$ the specific enthalpy. Equation \ref{eq:rs2} only depends on the jet power, radius and velocity for $h_{\rm j} \simeq c^2$ (i.e., for a cold jet). The angular size that such a value of $R_s$  would imply is even below the micro-arcsecond for sources at $\simeq 20\,{\rm Mpc}$. Therefore, it is very difficult to observe these interactions, unless the stellar wind is very powerful, with mass-loss orders of magnitude larger than $\dot{M}_{\rm w}\, =\, 10^{-11}\,{\rm M_\odot \, yr^{-1}}$ and in nearby AGN jets, as it has been suggested in the case of Centaurus A \cite{wo08,go10}. In \cite{br12,dlc16,pe17b}, the authors have studied the detail of jet-star/cloud interactions in the RHD approximation. They show that the local dissipation of jet kinetic energy in the interaction region can be efficient and that mixing can take place rapidly by means of the development of helical instabilities in the shocked wind tail. 
  
  Several works have pointed out the possible relevance of such interactions in the production of X-rays along the jet (e.g., \cite{wyk13,wyk15,vi17}), and also gamma-rays close to active galactic nuclei (e.g., \cite{bp97,bba10,ara13,tab19}). The particle acceleration is modelled to take place within the interaction region (via the Fermi I process, mainly) and the gamma-rays typically produced by inverse Compton scattering of stellar photons, or an external photon field if the interaction takes place close to the active nucleus. 
  
   Downstream of the obstacle, a cometary tail of shocked gas forms, which is surrounded by shocked jet plasma. The expansion of this over-pressured gas with respect to its environment favours acceleration along the tail. Furthermore, depending on the conditions, the tail can be easily destabilised and trigger mixing \cite{pe17b}. The global dynamical role of mass-loading by stars and other obstacles can be studied via relativistic magnetohydrodynamical (RMHD) simulations, by introducing a source term in the mass equation (the particles being injected with zero velocity and temperature relative to the jet) for each numerical cell, accounting for the number of stars per unit volume and distance to the nucleus. Then, assuming steady-state conditions, we can isolate the role of entrainment \cite{bo96}, or, by running dynamical simulations, we can test the combined effect of entrainment plus other processes, such as the development of instabilities \cite{pe14}.
   
   Steady-state simulations \cite{bo96} (see also Perucho et al., in preparation) already showed that this represents an efficient global mechanism for energy dissipation in jets. In \cite{bo96}, the authors reported that cold jets could be globally heated by the dissipation induced by the injection of mass by a distribution of stars. In terms of radiative output, this process would trigger strong emissivity \cite{wyk13,wyk15,vi17}, just as observed in the decelerating regions of FRIs \cite{lb14}. However, the modelling of those regions reveals a progressive deceleration from the jet boundaries. This means that, although mass-load by stars may efficiently contribute to jet deceleration, for the case of relatively weak jets ($L_j \sim 10^{42}\,{\rm erg/s}$) and old populations of low-mass stars ($\dot{M}_w \sim 10^{-12} M_\odot \,{\rm yr^{-1}}$), or for more powerful jets ($L_j \sim 10^{43}\,{\rm erg/s}$) and a relatively large number of red giants ($\dot{M}_w \sim 10^{-9} M_\odot \,{\rm yr^{-1}}$), it still does not seem to be the only answer to the dichotomy. Only if the bubbles formed by the shocked stellar winds in the interstellar medium are rapidly eroded at the jet shear-layer and they largely contribute to entrainment could the model be reconciled with the observations: In that situation, most of the entrainment would occur at the jet boundaries, and it would be dragged inwards by turbulent mixing. However, there are doubts about this \cite{tab19}. 
         
   Dynamical simulations \cite{pe14} show that the energy dissipation and heating of the jet induce expansion, which favours the penetration of more stars and, via deceleration, it also increases the growth-rates of different modes. In the case of axisymmetric simulations, only pinching (symmetric) modes can develop and the jets are disrupted by a strong shock, once the pinch becomes non-linear. This is an interesting example on how different processes can act together: In such a case, it would not make much sense to attribute jet deceleration to one or the other, but it should be attributed to the combination of both. 
  
  Particle acceleration can, in the scenario of jet-star interactions, take place both locally at shocks, and via extended acceleration in the turbulent mixing tails. It is certainly difficult to observe a single interaction (Eq.~\ref{eq:rs2}), but, owing to the large amount of stars embedded in the jet at a given time, it could certainly contribute to diffuse emission such as that detected in X-rays along the deceleration region \cite{lb14,kh12} in FRI jets.  Another relevant role of mass entrainment, in terms of radiative output, is the reduction of Doppler boosting, which makes aligned, distant jets, more difficult to detect, and has the opposite effect in misaligned, closer jets.  

   Assuming that the magnetic flux is very small compared to the kinetic flux beyond the jet flow acceleration zone, one can estimate the distance to which the jet momentum is completely consumed by the acceleration of entrained particles by equating the jet initial momentum of the jet  ($L_j/(\gamma_j\,c$, with $v_j \simeq c$) to the entrained mass \cite{hb06}:  
  
\begin{eqnarray} \label{eq:hb}
l_{\rm d} \simeq \frac{1} {\gamma_{\rm j}} \! \! \left(\frac{L_{\rm j}} {10^{43}\, {\rm erg\,s}^{-1}}\right)
\!\! \left(\frac{\dot{M}} {10^{-11}\, {\rm M_\odot} {\rm yr}^{-1}}\right)^{-1} \left(\frac{n_s} {1\,{\rm pc}^{-3}}\right)^{-1}\,  \!\!
\! \left(\frac{R_{\rm j}} {10\, {\rm pc}}\right)^{-2} \!\! \, 10^2\, {\rm kpc},
\end{eqnarray}
where $\dot{M}$ is the mean mass-loss rate of the stellar population in the galaxy, and $n_s$ is the number of stars per unit volume. From this expression, we see that a large contribution from stars is needed to completely decelerate a FRI jet with $L_{\rm j} = 10^{43} {\rm erg\,s^{-1}}$ within 1~kpc. This can only happen if the stellar population is dominated by red giants. However, because $l_d$ is inversely proportional to the square of the jet radius, and taking into account that energy dissipation produces an increase of the jet pressure (and thus, enhances expansion and favours the penetration of more stars), the deceleration process can undergo a certain degree of feedback, shortening $l_d$. Furthermore, there are other aspects that have to be taken into account when considering Eq.~\ref{eq:hb}: 1) The jet momentum is never completely depleted in FRI jets, so $l_d$ is, in this respect, an upper limit of the distance in which the jet becomes transonic, and 2) the number of stars drops with distance, and this would increase the value of $l_d$.

\section{Discussion}

\subsection{Entrainment and dissipation}

  Equation~\ref{eq:hb} provides an estimate of the distance at which the jet's momentum is completely absorbed by the entrained particles and the flow is thus stopped. We can make another estimate about the distance at which the mass load starts to be relevant with respect to the injected mass flux. Comparing the mass flux at two different positions along the jet, we obtain:
  
   \begin{equation} \label{eq:mcons}
  \rho_{\rm j}\, \gamma_{\rm j} \,R_{\rm j}^2 \,v\, = \,\rho_{\rm j,0} \, \gamma_{\rm j,0} \, R_{\rm j, 0}^2\,  c\, +\, \dot{M} \,n_s \, R_{\rm j}^2\, \Delta z, 
  \end{equation}
where the subscript $0$ indicates an initial location where the mass entrainment is negligible, and we assume $v_{\rm j,0} \simeq c$. Taking a constant opening angle, we can write $R_{\rm j} \simeq  \Delta z\,  \tan(\alpha_{\rm j})$, for large enough values of $\Delta z$. Then, comparing the two terms on the right-hand-side, we can state that mass-load will start to be relevant when
    \begin{equation} \label{eq:mcmp}
  \dot{M} \, n_s \, \tan(\alpha_{\rm j})^2 \,(\Delta z)^3 \, \simeq \, \rho_{\rm j,0}\, \gamma_{\rm j,0} \,R_{\rm j, 0}^2 \,c.  
  \end{equation}
  
  We can also write the initial mass flux in terms of the jet power:
 \begin{equation} \label{eq:mpow} 
  \rho_{\rm j,0} \,\gamma_{\rm j,0}\, \pi \,R_{\rm j,0}^2 \,c\, \simeq\, \frac{L_{\rm j}}{\gamma_{\rm j,0} \,c^2},
  \end{equation}    
 which is valid when $v_{\rm j,0} \simeq c$ and specific enthalpy is $h_{\rm j} \simeq c^2$ (i.e., cold jet). We derive, substituting $\Delta z$ by $l_m$, and $R_{\rm j}$ by $\Delta z \, \tan(\alpha_{\rm_j})$ (half opening angle): 
 
\begin{eqnarray} \label{eq:l_m}
l_m \simeq 390 \left( \frac{1} {\gamma_{\rm j,0}\,(\tan(\alpha_{\rm j}))^2}\, \! \! \left(\frac{L_{\rm j}} {10^{43}\, {\rm erg\,s}^{-1}}\right)
\!\! \left(\frac{\dot{M}} {10^{-11}\, {\rm M_\odot} {\rm yr}^{-1}}\right)^{-1} \left(\frac{n_s} {0.1\,{\rm pc}^{-3}}\right)^{-1}\, 
\! \right)^{1/3}  \!\! \, {\rm pc}.
  \end{eqnarray}     
This expression is not very sensitive to changes of the parameters in one order of magnitude, and reveals the possible relevant contribution to deceleration of stellar populations in which the number of red giants is large enough to increase the mean mass-loss rates. Again, the expression does not take into account the drop in $n_s$ with distance, but a large drop is not expected within the inner kiloparsecs, mainly in the giant ellipticals typically hosting FRI jets. For a small opening angle of $1^\circ$, we obtain $l_m\simeq 8.4/\gamma_{\rm j,0}\,{\rm kpc}$. An increase of the opening angle could compensate for the drop in the number of stars with distance: Actually, an opening angle of $1.5^\circ$ would reduce the estimate of the $l_m$ to 1~kpc. For a jet in free expansion, $\alpha_{\rm j} \simeq 1/\gamma_{\rm j}$, so $\alpha_{\rm j}\simeq 5.7^\circ$  if $\gamma_{\rm j}\simeq 10$. In this case, $l_m\leq 200\,{\rm pc}$. Interestingly, the distance obtained is within the order of magnitude of the expected deceleration scales in FRIs. 

   If the jet energy flux is dominated by the rest-mass of the particles, this distance will coincide with the deceleration distance. However, this is probably not the case at parsec scales, where the Lorentz factor can be large, and the internal or magnetic energy can be relevant. In that case, $l_m$ indicates the distance at which the jet particle composition is significantly changed. The relative importance of the energy flux associated to the mass of the particles will determine to which extent this is also dynamically relevant. 
   
   As a consequence of the previous statement, there are two options to interpret the situation in FRI jets, where deceleration takes place at hundreds of parsecs to a few kiloparsecs: 1) if this is caused by stars mainly, it means that the jet should be particle dominated at $z\simeq 100$~pc, and 2) if the jet energy flux is dominated by the magnetic or internal (hot jet) energy flux or it has a large Lorentz factor, then it becomes difficult to explain the observed behaviour only with stellar mass-load, unless the population of red giants or young stars is important enough.
 
    Beyond the deceleration distance, the jet becomes transonic and particle acceleration and significant radiative output will mainly take place in the turbulent regions triggered by entrainment (see \cite{bi84,bi86a,bi86b,dy86,ko88,ko90a,ko90b}). It is also important to stress that, although the above expressions are derived for stellar mass-loss, the terms $(\dot{M} \,n_s)^{-1}$ can be substituted by the entrainment from the jet boundaries, although in this case the modelling should take into account the inhomogeneity of the loading, which takes place radially inwards from the boundaries (see, e.g., \cite{wa09}).

It has been reported that the gamma-ray observatory \emph{Fermi} detects relatively more FRIs than FRIIs (see \cite{dg16,gr12}), and this cannot be reproduced in terms of population numbers. A straight temptation to explain this fact is to claim that dissipation of kinetic energy is stronger in FRIs. However, even if the fraction of dissipated energy in FRIIs is smaller, their larger energy budget could compensate that difference in the investment of kinetic energy into particle acceleration. Therefore, extra causes are needed. One possible option is Doppler boosting: while the gamma-ray emission is deboosted in FRIIs because of misalignment with our line of sight, the deceleration of FRI jets could favour detections at large viewing angles. This possibility would be supported by the detection of FRIIs with longer integration times. Another option is that the processes taking place in FRIs are more efficient in terms of particle acceleration than those in FRIIs. In this case, the main different mechanism is the turbulence triggered by entrainment (either from ambient gas or from stellar winds). As we have seen, we expect dissipation in FRIs to be caused by interaction with obstacles (shocks plus turbulence) within the inner kiloparsecs and entrainment produced by small-scale instabilities, and by turbulent mixing beyond this region. Then, the different detection fractions would demand such process to be a more efficient particle accelerator than shocks. The detection of extended emission in a nearby FRI radiogalaxy (Centaurus A, \cite{ab10}) indicates that, indeed, turbulence could be an efficient accelerator.

\subsection{A note on Microquasars}

    Although microquasar jets may form via the same physical mechanisms as extragalactic jets and may resemble them at large scales \cite{mar17}, the environment through which they evolve is extremely different from that through which extragalactic jets propagate. In the case of young microquasars, the jet may cross the companion wind, shocked wind, the supernova remnant (SNR), and the shocked interstellar medium (ISM), before entering the ISM. In the case of an aged microquasar, the jet has to get through the wind, shocked-wind, and shocked ISM \cite{brpb12}.
  
    Once triggered and accelerated, the jet suffers the lateral impact of the companion's wind \cite{pbr08,pbrk10,brb16,yoo16}. Depending on the wind-to-jet momentum ratio, oblique shocks and bends can be developed in the jet. The inhomogeneities in the winds of massive companions, which may embed dense clumps, can play the role of clouds in extragalactic jets, enhancing mass-loading and energy dissipation \cite{pbr12}. 
    
    In this initial region, shocks and non-linear effects on the jet should be expected. Therefore, strong kinetic and magnetic energy dissipation must take place. Beyond the binary region, the jet radius of the generated outflow is of the order of the size of the orbit, so that, unless the companion's wind disrupts it, the outflow is the result of the expansion of the injected plasma from the compact object and the bending/entrainment produced by the wind.      
      
   When the jet crosses any of the discontinuities defined by the SN or wind shocks, it enters denser media. In this case, dissipation must also occur at the strong interaction, which would show up as a temporary structure, before the jet carves its way through the shocked gas. Transient jet head deceleration is also expected. Finally, once the jet opens its way up to the ISM, it evolves through a fairly homogeneous medium \cite{bo+09}. 
   
   As we can conclude from the previous paragraphs, the evolution of microquasar jets takes place in environments that are completely diverse from those found by extragalactic jets. The latter are typically formed from non-orbiting black-holes (but in some specific cases, probably), and find inhomogeneous media in the inner kiloparsecs, as they cross the ISM, but evolve, grossly speaking, through decreasing density and pressure environments. On the contrary, the former are formed along orbits that are much wider than their injection radii, suffer the impact of a lateral wind and cross several discontinuities into denser media. These all represent non-linear processes,\footnote{In powerful jets and low mass companions, the impact of the wind could be taken as a linear perturbation.} which can drastically change the jet evolution and trigger an efficient acceleration of particles.

\section{Conclusions}

Altogether, the large amounts of energy transported by jets and a series of dissipative processes like the ones described here (or others, as magnetic reconnection in small scales), give a reasonable scenario to explain the high and very-high energy emitted from jets. The growth of different instability types and the interaction of the jet flow with different types of obstacles probably represent the two main scenarios by which jets dissipate energy, via shocks, shear and turbulence along the inner kiloparsecs of jet evolution. In the case of large scales FRI jets, once the flows become transonic, the most probable process to accelerate particles is turbulence, 

  Although the evolution of microquasar jets must be very different to that of extragalactic jets, due to the different environmental conditions, analogous non-linear processes can take place in this case, such as interactions with dense inhomogeneous clumps formed in the stellar winds of massive companion stars, or helical instabilities induced by the orbital motion of the injection point, or the stellar wind. 

A difficult problem to overcome from a theoretical perspective is the link between these large-scale plasma scenarios and the detailed particle acceleration process, mainly because of the orders of magnitude difference in spatial scales. Overcoming this gap is a big challenge that numerical techniques will have to face for the exact relation between the macroscopic phenomena and the radiative output from relativistic outflows to be understood.

\section*{Acknowledgements}

I thank the SOC and LOC of the meeting for their hospitality and the organization of a very interesting meeting. I also thank the referee of this contribution for his/her positive comments. MP has been supported by the Spanish Ministerio de Ciencia y
Universidades (grants AYA2015-66899-C2-1-P and AYA2016-77237-C3-3-P)
and the Generalitat Valenciana (grant PROMETEOII/2014/069).

\end{document}